\documentclass{nature}
\usepackage{graphicx}
\usepackage{pdflscape}
\usepackage{hyperref}
\usepackage{txfonts}
\usepackage{textcomp}
\usepackage{threeparttable}
\usepackage{xcolor}
\usepackage{caption}
\usepackage[caption = false]{subfig}

\makeatletter
\renewcommand{\baselinestretch}{1.2} 

\let\saved@includegraphics\includegraphics
\AtBeginDocument{\let\includegraphics\saved@includegraphics}
\renewenvironment*{figure}{\@float{figure}}{\end@float}
\makeatother

\newcommand{\D}{\mathrm{d}}

\newcommand{\EQ}[1] {equation~(\ref{#1})}
\newcommand{\FIG}[1] {Fig.~\ref{#1}}
\newcommand{\TAB}[1] {Table~\ref{#1}}
\newcommand{\EXTTAB}[1] {Extended Data Table~\ref{#1}}
\newcommand{\EXTFIG}[1] {Extended Data Fig.~\ref{#1}}

\def\DM{\mathrm{DM}}

\def\DMd{\DM_{\rm halo}}

\def\cmpc{\mathrm{cm}^{-3}\, \mathrm{pc}}
\def\radm{\mathrm{rad}\,\mathrm{m}^{-2}}

\definecolor{dkblue}{RGB}{54, 86, 169}
\title{Diverse polarization angle swings from a 
repeating fast radio burst source}

\author{R. Luo$^{1,2,3}$, B. J. Wang$^{1,2}$, Y. P. Men$^{1,2}$, C. F. Zhang$^{1,2}$, J. C. Jiang$^{1,2}$, H. Xu$^{1,2}$, W. Y. Wang$^{2,4}$, 
K. J. Lee$^{1,2}$\thanks{E-mail: kjlee@pku.edu.cn,\href{https://orcid.org/0000-0002-1435-0883}{\hspace{2mm} orcid.org/0000-0002-1435-0883}},
J. L. Han$^{2,4,5}$\thanks{Email: hjl@nao.cas.cn,\href{https://orcid.org/0000-0002-9274-3092}{\hspace{2mm} orcid.org/0000-0002-9274-3092}},
B. Zhang$^{6}$\thanks{Email: zhang@physics.unlv.edu, \href{https://orcid.org/0000-0002-9725-2524}{\hspace{2mm} orcid.org/0000-0002-9725-2524}},
R. N. Caballero$^{1}$, M. Z. Chen$^{7}$, X. L. Chen$^{2,4}$, H. Q. Gan$^{2,5}$, Y. J. Guo$^{1,8}$, L. F. Hao$^{9}$, Y. X. Huang$^{9}$, P. Jiang$^{2,5}$, H. Li$^{2,5}$, J. Li$^{7}$,  Z. X. Li$^{9}$, J. T. Luo$^{10}$, J. Pan$^{2}$, X. Pei$^{5,7}$, L. Qian$^{2,5}$, J. H. Sun$^{2,5}$, M. Wang$^9$, N. Wang$^7$, Z. G. Wen$^{7}$, R. X. Xu$^{1,11}$, Y. H. Xu$^{9}$, J. Yan$^{2,5}$, W. M. Yan$^7$, D. J. Yu$^{2,5}$, J. P. Yuan$^{7}$, S. B. Zhang$^{3,4,12}$ \& Y. Zhu$^{2,5}$}

\begin{document}
\maketitle
\begin{affiliations}
 \item Kavli Institute for Astronomy and Astrophysics, Peking University, Beijing 100871, China
 \item National Astronomical Observatories, Chinese Academy of Sciences, Beijing 100101, China
 \item CSIRO Astronomy and Space Science, Australia Telescope National Facility, Box 76, Epping, NSW 1710, Australia
 \item School of Astronomy, University of Chinese Academy of Sciences, Beijing 100049, China
 \item  CAS Key Laboratory of FAST, National Astronomical Observatories, Chinese Academy of Sciences, Beijing 100101, China
 \item Department of Physics and Astronomy, University of Nevada, Las Vegas, NV 89154, USA
 \item Xinjiang Astronomical Observatory, Chinese Academy of Sciences, 150 Science 1-Street, Urumqi 830011, China
 \item Max-Planck institut f\"ur Radioastronomie, Auf Dem H\"ugel, Bonn, 53121, Germany
 \item Yunnan Observatories, Chinese Academy of Sciences, Kunming 650216, China
 \item National Time Service Center, Chinese Academy Of Sciences, Xi'an 710600, China
 \item State Key Laboratory of Nuclear Physics and Technology, School of Physics, Peking University, Beijing 100871, China
 \item Purple Mountain Observatory, Chinese Academy of Sciences, Nanjing 210008, China
\end{affiliations}

\begin{abstract} 
Fast radio bursts (FRBs) are mysterious millisecond-duration radio transients\cite{Lorimer07Sci,Petroff19AAR}. Two possible mechanisms that could generate extremely coherent emission from FRBs invoke neutron star magnetospheres\cite{Kumar17MN,Zhang17ApJL,YangZhang18ApJ} or relativistic shocks far from the central energy source\cite{Lyubarsky14MN,Metzger19MN,Beloborodov19}. Detailed polarization observations may help us to understand the emission mechanism. However, the available FRB polarization data have been perplexing, because they show a host of polarimetric properties, including either a constant polarization angle during each burst for some repeaters\cite{Michilli18Nat,CHIME19ApJL}, or variable polarization angles in some other apparently one-off events\cite{Masui15Nat,Cho20}. Here we report observations of 15 bursts from FRB 180301 and find various polarization angle swings in seven of them. The diversity of the polarization angle features of these bursts is consistent with a magnetospheric origin of the radio emission, and disfavours the radiation models invoking relativistic shocks. 
\end{abstract}

Thanks to its high sensitivity, the Five-hundred-meter Aperture Spherical radio Telescope (FAST)\cite{Jiang19SCPMA} is well positioned to search for repeating bursts from FRB sources and carry out polarization observations. 
FRB 180301 was discovered by the Parkes 64-m radio telescope\cite{Price19MN}, and has a dispersion measure ($\DM$) of $522\,\cmpc$. We carried out four observations for this source in July, September and October 2019 with a total observing time of 12 h (see 
\EXTTAB{tab:obs}) using the 19-beam receiver mounted on FAST, covering 
the frequency range of $1000-1500$\,MHz. In the July and September sessions, the 
center beam (with a $3'$ beam size at full-width-half-maximum, FWHM) of the 19-beam receiver was pointed at the 
previously reported position (right ascension, $\alpha = 
06^{\mathrm{h}}12^{\mathrm{m}}43.4^{\mathrm{s}}$; declination, $\delta =  
+04^\circ33'45.4''$) from the Parkes discovery observation\cite{Price19MN} (where the telescope's FWHM beam size was $14.1'$). Four bursts were detected using Beam 7 in the July session, where we only recorded two linear polarization channels. Based on the fact that we did not detect any burst signal in the beams other than Beam 7, we estimated the FRB position to be at $\alpha = 06^{\mathrm{h}}12^{\mathrm{m}}54.96^{\mathrm{s}}$ and $\delta =  
+04^\circ38'43.6''$ with an error circle of $2.6'$ (see Methods). 
No burst above the detection threshold (signal-to-noise ratio, $\rm S/N\ge 8$) was found in 
the September session. In the two sessions in October, the FAST center beam was pointed to the July position of Beam 7, and 11 more bursts were detected, with the full polarimetric data recorded. The pulse profiles and dynamic spectra of these bursts with time and frequency are presented in \EXTFIG{fig:all_bursts} (see Methods). When they are dedispersed to maximize the frequency-integrated temporal structure, these bursts have DM$\simeq517\,\cmpc$ (see
\TAB{tab:rpts}). Their peak flux densities
range from 5.3 mJy to 94.1 mJy. 

The bursts show rich pulse structures without a common pattern. 
All the bursts have intrinsic widths of a few milliseconds with no apparent evidence
of scattering (see \TAB{tab:rpts}). These
features resemble those commonly seen in repeating
FRBs\cite{Spitler16Nat, CHIME19Nat02, CHIME19ApJL, CHIME20ApJL}.
Some bursts (e.g. burst 5) show a clear frequency down-drifting pattern, 
as also observed in some other repeating bursts\cite{Hessels19ApJL,
CHIME19Nat02, CHIME19ApJL, CHIME20ApJL}, suggesting a common physical origin.

\begin{figure} 
\centering
\includegraphics[width=\textwidth]{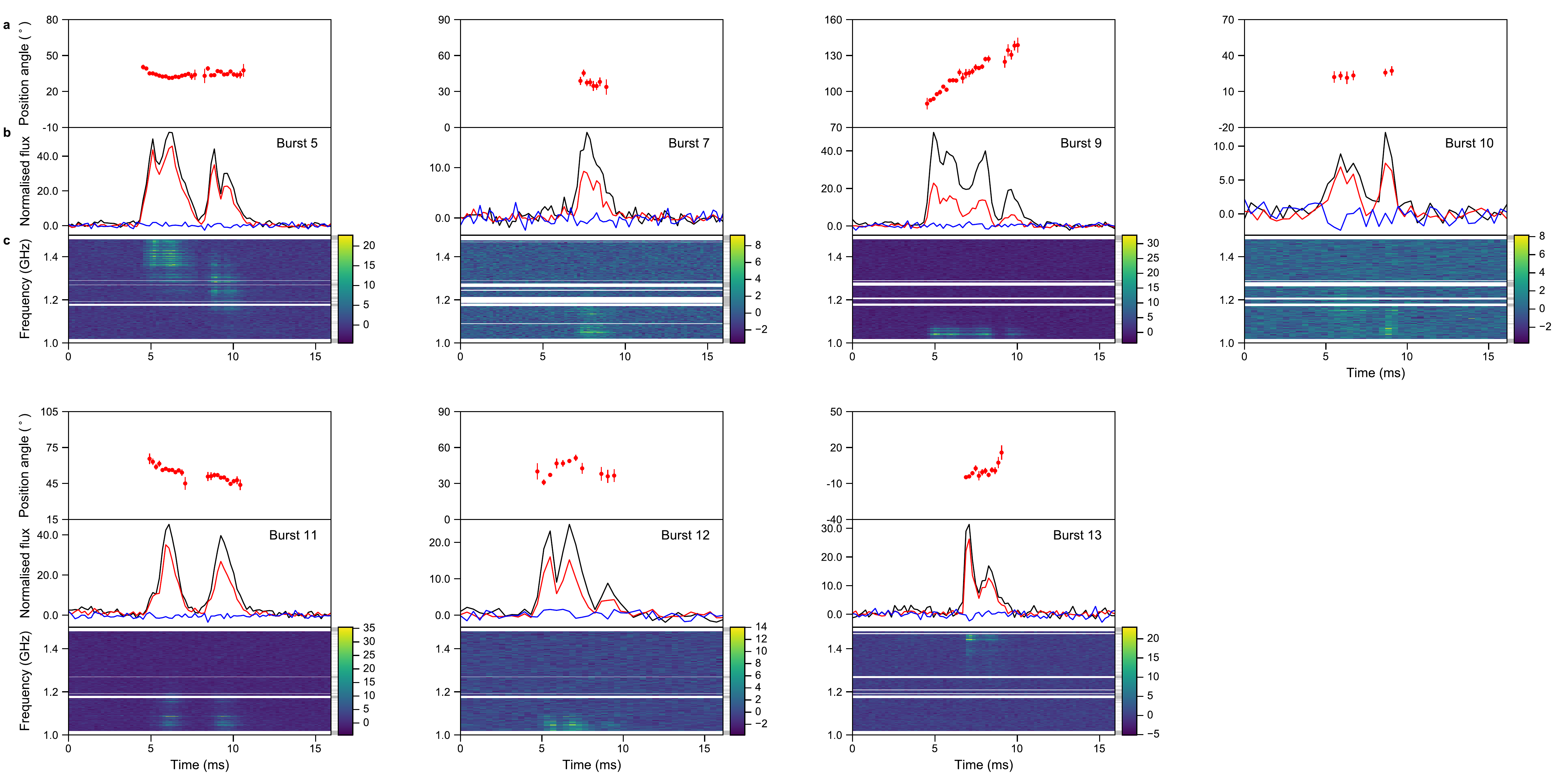}
\caption{\textbf{Polarization profiles and dynamic 
spectra of the seven brightest bursts from FRB 180301.} 
The Bayesian RM of each burst listed in \TAB{tab:rpts} is used 
to derotate the linear PA. Top, PA of linear polarization at infinite 
frequency; error bars represent 68\% confidence. Middle, polarization pulse 
profile; black, red and blue curves denote total intensity, linear 
polarization and circular polarization, respectively. Bottom, dynamic 
spectra for the total intensity as a function of frequency and time
(frequency resolution $1.95\,\mathrm{MHz}$ per channel and time resolution $196.6\,\mathrm{\mu s}$ per bin for all bursts except bursts 10 and 12, which are plotted with a time resolution of $393.2\,\mathrm{\mu}$ per bin). The colour bars denote the intensity S/N, that is, the flux scaled with the off-pulse r.m.s. amplitude. $\DM_{\rm aligned}$ in \TAB{tab:rpts} is used to dedisperse each burst. The PA curves are plotted in the same scale with a total range of 90$^\circ$. One can see a rich diversity of PA swings across the pulse profiles$-$for example, sweeping up for bursts 9 and 13, sweeping down for burst 11, or up and down for burst 12$-$even though some (for example burst 10) are consistent with a non-varying PA during the burst. 
\label{fig:poln2}}
\end{figure}

The bursts show a diversity of polarimetric
properties. Polarization pulse profiles of seven bursts with the best S/N polarization data, together with their dynamic spectra are presented in \FIG{fig:poln2}. These bursts have a high degree of linear polarization, some with a 
degree of polarization of $\Pi>70\%$. On the other hand, no obvious circular polarization was detected for any of the bursts, with an upper limit of 3\% for the strong bursts 5, 9, 11, 12, 13, and of about 10\% for weak bursts 7 and 10. These properties are similar to those of some repeating FRBs, such as FRB 121102, which show a nearly 100\% linear polarization and circular polarization of less than a few per cent \cite{Michilli18Nat}. However, FRB 180301 shows a rich diversity of  polarization position angle (PA) swings across the pulse profiles$-$for example, sweeping up for bursts 9 and 13, sweeping down for burst 11, or up and down for burst 12$-$although some (for example, burst 10) are consistent with a non-varying PA during the burst. These polarization features have not been detected in previous observations, which showed, for example, a nearly constant PA during each burst from repeaters such as FRB 121102\cite{Michilli18Nat} and FRB 180916.J0158+65\cite{CHIME19ApJL},
PA swings for some apparently non-repeating FRBs\cite{Masui15Nat,Cho20}, or the considerable amount of circular polarization from some FRBs\cite{Petroff17MN}. Our observations revealed a high level of diversity for PA swings of repeating bursts from one source, suggesting that polarimetric properties are not a good indication for distinguishing repeaters from apparent non-repeaters.

\begin{figure}
\centering
\includegraphics[width=\linewidth]{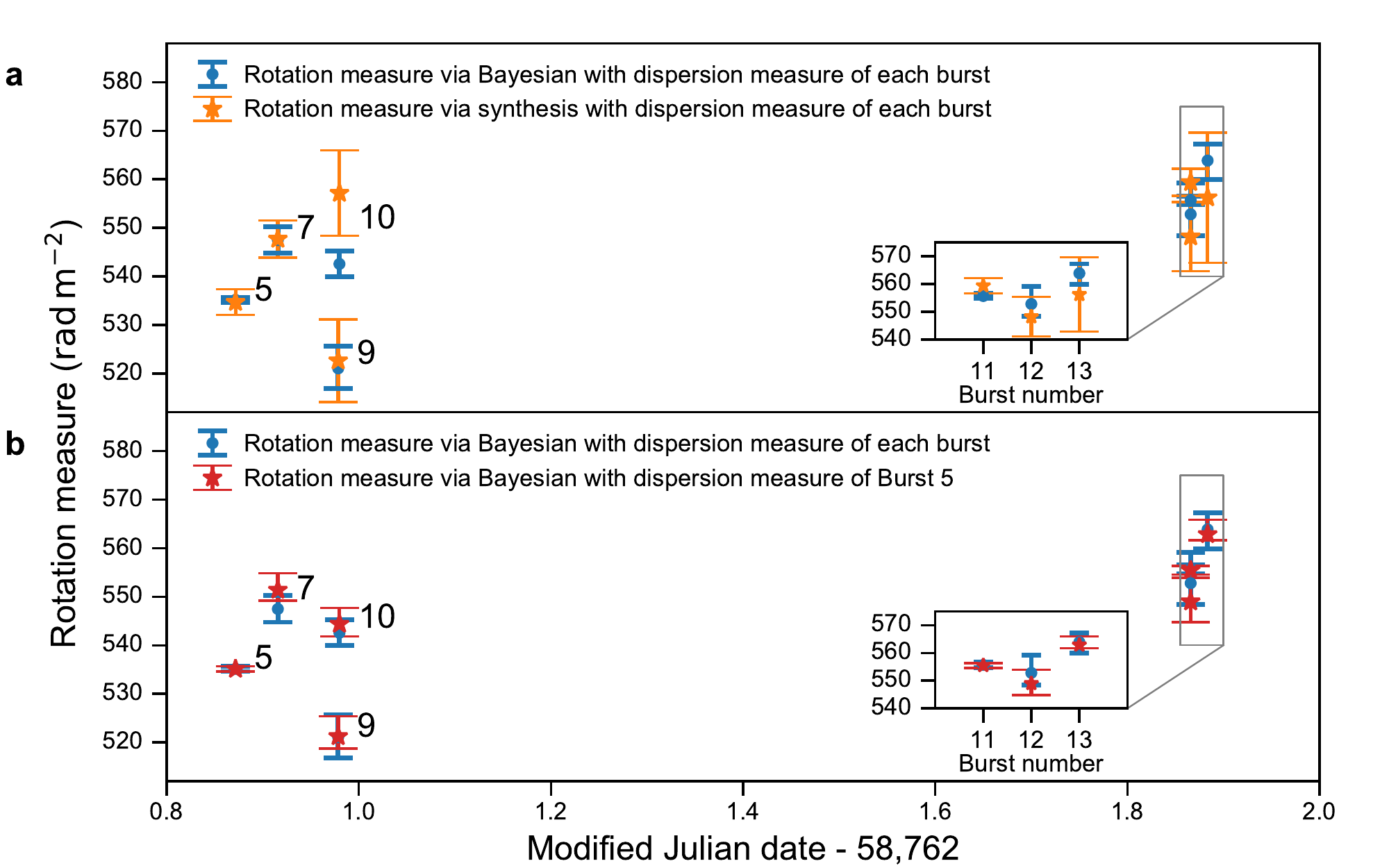}
\caption{{\textbf{RM values of seven bursts from FRB 180301.}  {\bf a}, Comparison between RM measured with the Bayesian and the RM synthesis method. The burst numbers are defined in \TAB{tab:rpts}. The error bars denote the 68\% confidence intervals. Blue points denote RMs computed using the Bayesian method with the $\DM_{\rm align}$ values in \TAB{tab:rpts}; orange stars represent RM values computed with the RM synthesis method. {\bf b}, Comparison between RMs calculated using $\DM_{\rm align}$ (blue points) and the DM of burst 5 (red stars). The details of RM inference are given in Methods.}
\label{fig:rmvar}}
\end{figure}

The detection of linear polarization allowed the measurement of the
Faraday rotation measure (RM; see details in Methods). For seven
bursts with polarization data recorded and $\rm S/N>5$ for the polarization intensity, we calibrated the
data and found similar RM values ranging from $521.5^{+4.6}_{-4.2}\,\radm$
to $564.4^{+3.4}_{-4.0}\,\radm$ (with 68\% confidence level), very different from that reported (${-3163\pm 20\,\radm}$) in the initial discovery\cite{Price19MN}. It is possible that the previous reported RM value was biased oweing to the narrow signal bandwidth of 40 MHz (see Methods). RM variations on a wide range of timescales have been reported for FRB 121102\cite{Michilli18Nat}, including a variation of about 0.5\% ($\sim 500\,\radm$) over a month and 10\% variation ($\sim10^{4}\,\radm$) over seven months. A small burst-to-burst RM change for FRB 121102 ($\sim50\,\radm$) cannot be ruled out. For FRB 180301, a $\chi^2$ test (see Methods) shows that the RM is unlikely a constant, varying on the timescale of one day as shown in \FIG{fig:rmvar}. Such an RM variation is not affected by the DM uncertainties, because the results are nearly the same when using either the DM values inferred from each burst or the DM of the burst with the highest S/N (burst 5).
Our measured RM values have a possible linear trend with a slope of
21$\,\radm\,{\rm d^{-1}}$ and {pulse-to-pulse} RM variations with a root-mean-square (r.m.s.) amplitude of 14$\,\radm$. 

FRB models typically address
both the source of the burst and the radiation mechanism. The polarization dataset presented here offers clues in understanding the radiation mechanisms of repeating FRBs.  
Two groups of radiation models have been widely used to interpret the 
extremely high brightness temperatures of FRBs. One group of models invokes a 
magnetosphere of a neutron star to radiate coherent emission\cite{Kumar17MN, 
YangZhang18ApJ}, whereas the other group invokes the synchrotron maser 
mechanism in relativistic shocks from a highly magnetized neutron star\cite{Metzger19MN, 
Beloborodov19}. Both groups of models are able to interpret strong linear 
polarization with non-varying PAs, as observed in FRB 121102 and FRB 180916.J0158+65.
The magnetospheric models invoke ordered field lines that sweep the line of sight as the source rotates\cite{RadhakrishnanApL1969} (as frequently observed in radio pulsars and flaring magnetars\cite{Levin12MN, Eatough13Nat, Camilo16ApJ}) or as the magnetospheric configuration is reshaped by an external ram pressure\cite{Zhang17ApJL}. Depending on the magnetospheric configuration and line-of-sight geometry, the magnetosphere models can produce diverse PA swing patterns (including a non-varying PA pattern\cite{Zhang18ApJL}). However, the synchrotron maser models require the ordered magnetic field in the shock plane to reach the coherent condition for radio emission. These models predict a constant PA during each pulse\cite{Metzger19MN,Beloborodov19} and have great difficulties in producing diverse PA variation patterns across pulses.
The observations reported here therefore
disfavour the synchrotron maser mechanism 
but support a magnetospheric origin of FRB radio emission (Methods).

The burst rate of FRB~180301 is $1.2^{+0.8}_{-0.7}\,\rm h^{-1}$ with a shape parameter $k=0.9\pm 0.3$ (see Methods) assuming a Weibull distribution\cite{Oppermann18MN}, when we count 15 bursts detected in the four sessions by FAST with a flux density threshold of 5 mJy. The shape parameter is close to 1, which suggests that the burst arrival times can be well described by a Poissonian distribution. The high burst rate makes FRB~180301 an active repeater, similar to FRB 121102.
FRB 121102\cite{Michilli18Nat} has an exceptionally large RM value, which has been interpreted as either due to a putative special 
location close to a supermassive black hole\cite{Michilli18Nat, Zhang18ApJL} or 
due to a strong magnetic field in the wind or supernova remnant of a very young 
magnetar\cite{Metzger19MN}. Within the magnetar scenario, if the burst rate is a proxy for the age of the underlying magnetar, other factors are needed to account for the RM discrepancy between FRB 180301 and FRB 121102. 

\clearpage

\clearpage

\captionsetup[table]{name={\bf Table}}
\captionsetup[figure]{name={\bf Fig.}}

\clearpage

\begin{landscape}  
\begin{table*}[htbp]
    \centering
    \caption{\textbf{Measured parameters of the bursts from FRB 180301}}
    \begin{threeparttable}
    \begin{tabular}{lcccccccccc}
    \hline
		 Burst number & Arrival Time\tnote{a} & $\DM_{\rm S/N}$\tnote{b} & $\DM_{\rm 
		 aligned}$\tnote{c} & $S$ $\tnote{d}$ & S/N & Width & $\tau_{\rm s}$\,\tnote{e}
		 & $\Pi$\,\tnote{f} & $\mathrm{RM_{Bayes}}$\,\tnote{g}& 
		 $\mathrm{RM_{syn}}$\,\tnote{h} \\ 
 & (MJD) & ($\cmpc$) & ($\cmpc$) & (mJy) & & (ms) & (ms) & (\%) & 
($\radm$) & ($\radm$) \\ 
    \hline
		1 & {58680.09189754} & 521.1$\pm$0.1 & 518.3$\pm$1.2 & 40.5 & 46.4 & 
		3.7$\pm$0.1 & {$<0.5$} & -- & -- & -- \\ 
		2 & {58680.11702228} & 522.3$\pm$1.4 & -- & 7.2 & 11.4 & 7.1$\pm$0.7 & -- & 
		-- & -- & -- \\ 
		3 & {58680.11785237} & 519.9$\pm$0.4 & -- & 11.5 & 11.5 & 2.8$\pm$0.3 & 
		$<0.1$ & -- & -- & -- \\ 
		4 & {58680.14004130} & 525.1$\pm$1.7 & -- & 8.6 & 12.9 & 6.3$\pm$0.5 & -- & -- & -- & -- \\ 
		5 & {58762.87166882} & 520.7$\pm$0.1 & 516.8$\pm$1.6 & 94.1 & 155.1 & 
		4.3$\pm$0.1 &{$<1$} & {$79.5\pm 1.0$} & 
		{$535.2_{-0.6}^{+0.5}$} & {$534.7\pm 2.6$} \\ 
		6\tnote{i} & {58762.90818370} & 517.3$\pm$0.7 & -- & 10.7 & 14.3 & 2.8$\pm$0.2 & 
		$<6$ & -- & -- & -- \\ 
		7 & {58762.91573324} & 517.3$\pm$0.2 & 516.6$\pm$0.7 & 18.8 & 29.5 & 
		3.9$\pm$0.1 & $<0.1$ & {$46.9\pm 4.5$} & 
		{$547.5_{-2.7}^{+2.8}$} & {$547.7\pm 3.8$} \\ 
		8\tnote{i} & {58762.94422687} & 510.3$\pm$2.3 & 516.6$\pm$1.3 & 6.8 & 14.4 & 
		7.1$\pm$0.6 & $<10$ & -- & -- & -- \\ 
		9 & {58762.97852743} & 517.1$\pm$0.2 & 516.3$\pm$2.6 & 60.2 & 120.1 & 
		6.3$\pm$0.1 & {$<4$} & $36.6\pm 0.9$ & 
		{$521.0_{-4.2}^{+4.6}$} & {$522.6\pm 8.5$} \\ 
		10 & {58762.97985277} & 518.9$\pm$0.3 & 517.0$\pm$0.6 & 10.3 & 17.7 & 
		4.7$\pm$0.3 & {$<0.7$} & {$67.4\pm 7.0$} & 
		{$542.6_{-2.6}^{+2.7}$} & {$557.2\pm 8.8$}  \\ 
		11 & {58763.86585273} & 517.1$\pm$0.1 & 516.6$\pm$0.8 & 43.2 & 88.9 & 
		6.7$\pm$0.9 & $<0.2$ & $69.0\pm 1.4$ & 
		{$555.7_{-0.9}^{+0.9}$}& {$559.4\pm 2.8$} \\ 
		12 & {58763.86585290} & 516.9$\pm$0.4 & 515.9$\pm$2.3 & 21.2 & 39.5 & 
		5.5$\pm$0.2 & {$<4$} & {$55.6\pm 2.9$} & 
		{$552.8_{-4.4}^{+6.4}$} & {$548.7\pm 7.1$} \\ 
		13 & {58763.88351268} & 516.9$\pm$0.2 & 516.7$\pm$1.7 & 32.5 & 45.5 & 
		3.1$\pm$0.1 & {$<10$} & {$73.6\pm 3.2$} & 
		{$563.9_{-4.0}^{+3.4}$} & {$552.7\pm 13.4$} \\ 
		14\tnote{i} & {58763.92197253} & 522.7$\pm$1.4 & -- & 7.1 & 17.7 & 9.8$\pm$0.6 & -- & 
		-- & -- & -- \\ 
		15\tnote{i} & {58763.97110374} & 525.5$\pm$0.6 & -- & 5.3 & 9.8 & 5.5$\pm$0.6 & -- & 
		-- & -- & -- \\ 
    \hline
    \end{tabular}
    \begin{tablenotes}
    \footnotesize
    
    \item a Infinite-frequency arrival times at the Solar System barycentre.
    \item b DM obtained by maximising S/N.
	\item c DM obtained by the best {burst} alignment.
	\item d Peak flux density. Uncertainties are dominated by the variation of the system noise temperature ($\sim20$\%)\cite{Jiang19SCPMA}.
    \item e Scattering timescale at 1 GHz at 68\% confidence level, determined by curve fitting\cite{SL14}.
	\item f Degree of polarization.  
	\item g RM at 68\% confidence level, obtained by the 
			Bayesian method. 
	\item h RM at 68\% confidence level, obtained by RM 
			synthesis. 
	\item i Because of the low S/N ($\le 5$), we neglect the degree of polarization for bursts 6, 8, 14, 15.
    \end{tablenotes} 
    \end{threeparttable}
    \label{tab:rpts}
    \end{table*}
\end{landscape}

\clearpage

\clearpage

\section*{Methods}

\subsubsection*{Observations and burst detection }

The observations were carried out using the FAST telescope which
has an effective collecting area of 300~m in diameter, mounted with
the 19-beam receiver. The system temperature is about $20-25$ K for
different beams, as reported in the specifications of the FAST L-band system\cite{Jiang20RAA}. The designed frequency response
of the 19-beam receiver covers from 1050 to 1450 MHz, with the
central frequency at 1250 MHz; in reality, signals of 1000
to 1500 MHz with a degraded bandpass of 50 MHz at each of the two edges are
also received and down-converted. The raw voltages were digitized
using a pulsar backend based on the Re-configurable Open Architecture
Computing Hardware-2 (ROACH2) system\cite{HA16}, where the voltage
data are correlated and integrated to form filterbank data with a
49.152 $\mu s$ time resolution. The entire 500-MHz bandwidth is
divided into 4096 channels. As a search observation, two
orthogonal linear polarization intensity channels were recorded.

Observations were carried out in four sessions (see \EXTTAB{tab:obs}).
During the first session on 16 July 2019, we detected four bursts over a 2-h observation, and the burst signal was only seen in Beam 7 (out of 19 beams). Later in the same year,  4-, 3- and 3-hour follow up
observations were carried out on 11 September, 6 October, and 7 October, respectively. No burst was detected in the second session, but six and five bursts with full polarization Stokes parameters were recorded in the 3rd and the 4th sessions when the central beam of the receiver was placed on the new position (see below).

We used the \textsc{BEAR}(Burst Emission Automatic Roger) package\cite{Men19MN} to perform the FRB searches. \textsc{BEAR} is 
capable of performing radio-frequency interference (RFI) mitigation, de-dispersion, 
and candidates score ranking in one run. When searching for bursts, data at 
$1200-1250$\,MHz were excluded to avoid satellite interference. We de-disperse the 
data with DM trials in 500 to 550\,$\cmpc$ in steps of 0.5\,$\cmpc$. To reduce the computational cost, the 
pulse width was searched from 0.2 ms to 20 ms in the box-car-shaped matched 
filter of \textsc{BEAR}. The signals were then sifted visually 
for candidates with $\rm S/N>8$. The RFI at the FAST site is dominated only 
by satellite signals thanks to the radio protection zone. The narrow-band, 
close-to-zero DM signals were classified as RFI. Although we 
searched only with a DM range of $500-550\,\cmpc$, the zero-DM time series 
was always computed by the RFI filter of \textsc{BEAR}.
Signals with a strong indication of cold-plasma dispersion are recorded as 
FRB candidates. We also plotted all the candidates using a graphical interface, 
where RFI and FRB signals can be well separated in the DM and 
S/N space. 
The dynamic spectra of all 15 bursts detected are presented in the 
\EXTFIG{fig:all_bursts}.

\subsubsection*{Localisation and {positional accuracy}}
All the 19 beams have an average full half-power
beam width (HPBW) of $3'$, and the beams were spaced by $5.74'$ in
a cellular structure \cite{Jiang20RAA}. Because burst 5$-$the strongest one 
that we detected, with $\mathrm{S/N}=155-$was not detected in the six beams 
surrounding the central beam above the S/N threshold of 8, the angular 
distance of FRB 180301 from the centre of Beam 1
cannot be larger than $\sim2.6'$. Otherwise, it would appear in
the sidelobes of other beams. The pointing error of FAST
is less than $15''$, which is much smaller than the HPBW. We therefore conclude that the position of this FRB is located at $\alpha =
06^{\mathrm{h}}12^{\mathrm{m}}54.96^{\mathrm{s}}$ and $\delta =
+04^\circ38'43.6''$ with a position uncertainty of $2.6'$ (radius, 68\% confidence level).

\subsubsection*{Dispersion measure}
DM values obtained by aligning the signal across frequencies to achieve the best peak
S/N(ref.\cite{LK12HPA}) or by maximizing the temporal structure will be different as noted in 
the literature\cite{Hessels19ApJL}. Here we report both kinds of DM values (\TAB{tab:rpts}). Deriving the DM from the maximization of S/N has been well documented\cite{LK12HPA}, and we refer the interested reader to the standard reference.
To maximize the temporal structure, we maximize the
S/N-weighted 'local contrast' of dedispersed pulse profile as defined in \EQ{eq:ktrst}. {Previously, this was done} 
without the S/N weight\cite{Hessels19ApJL}. The estimated $\widehat{\DM}$ is 
\begin{equation}
    \widehat{\DM}= {\rm argmax}_\DM \max\left\{\frac{ [s(\DM)_{i}-s(\DM)_{i-1}]^2}{s(\DM)_i+s(\DM)_{i-1}} \left[\frac{s(\DM)_i+s(\DM)_{i-1}}{\sigma} \right]^2\right\}\,, \forall i\in [2,N]\,,
    \label{eq:ktrst}
\end{equation}
for $N$ data points in the dedispersed pulse profile
($s$). We use the r.m.s value of the integrated profile baseline to 
estimate the profile error ($\sigma$). The first term in the equation is the 
local contrast, that is, the ratio between the difference of the $i$th bin and $(i-1)$th bin 
($s_i-s_{i-1}$) and their mean. The second term is the S/N weight.  In 
practice, we dedisperse the dynamic spectra in a series of DM trials to get 
the profiles. Then, we calculate the maximum of the local contrast of each profile to get the best alignment of the pulses. 
The optimal DM is the $\DM_{\rm aligned}$ in \TAB{tab:rpts}.  
The DM error is determined by the FWHM of the contrast-DM curve. 
This approach gives a better determined DM (weighted mean $\DM=516.76\,
\cmpc$ with standard deviation of $\Delta(\DM)=0.62\,\cmpc$) compared 
to the previously used method\cite{Hessels19ApJL} (weighted mean $
\DM=516.23\,\cmpc$ with standard deviation of $\Delta(\DM)=1.19\,\cmpc$), 
because the S/N weighting in our method helps to reduce the uncertainty.

\subsubsection*{Redshift inference}
The redshift of FRB~180301 is inferred using the method described by Luo et 
al.\cite{Luo18MN}, in which the cosmic co-moving volume size is used as the 
distance prior and galaxy synthesis is applied to determine the likelihood 
function of the observed DM with a certain redshift $z$. 
The extragalactic DM was obtained by removing the foreground contribution 
from the Milky Way using both the NE2001\cite{CL02arXiv} and the YMW16\cite{YMW16} models.  
We used the DM from the halo with the mean value $\DMd\simeq30\,\cmpc$ from the simulation\cite{Dolag15MN}, and the ${\rm 
DM}_{\rm IGM}-z$ relation is adopted using the standard formula\cite{Deng14} 
with the intergalactic medium fraction $f_{\rm IGM} \sim 0.83$. 
We inferred the redshift using the maximal posterior inference. 
According to the galaxy synthesis model\cite{Luo18MN}, the redshift at 95\% confidence level is $z=0.20-0.35$ 
using the NE2001 model and $z=0.13-0.23$ using the YMW16 model.

\subsubsection*{Polarization and Faraday rotation }
We performed the polarization calibration with the \textsc{psrchive}\cite{HvSM04} software package (\url{http://psrchive.sourceforge.net})
using the single-axial model, in which the differential gain and phase are calibrated using the noise diode.
Such a calibration scheme can reduce the systematics down to the
0.5\% level, as the leakage term is better than $-46$\,dB within the FWHM 
region of the central beam as measured in the FAST engineering 
phase\cite{DBC17}. The polarization data 
were recorded from $1000-1500$ MHz. The feed efficiency drops below 70\% and noise temperature rises steeply\cite{DBC17} near the band edge; 
therefore, we only used data in the range $1020-1480$\,MHz for later analysis. We checked the bandpass of Stokes intnsity parameter $V$ in the pulse region, as shown in \FIG{fig:poln2}. The fluctuation is compatible with the r.m.s. amplitude in the off-pulse region, which indicates that 
the differential phase between the two linear feeds was calibrated properly.

Among the 11 bursts detected in the third and fourth 
observations, seven have high S/N for the polarization intensity ($\rm S/N_{pol}\ge 5$), with which the Faraday rotation can be measured. {Frequency channels containing strong RFI were excluded}, as marked in \EXTFIG{fig:qu}. The bins with integrated S/N$<5$ were also excluded. 

There are two major classes of methods with which to measure the RM, 1) $Q-U$
fitting, which fits the Stokes intensity $Q$ and $U$ using sinusoidal
curves\cite{DKL19}; and 2) RM synthesis\cite{rmsyn15}, which derotates
the RM-induced PA wrap and uses the maximum of the total linear
polarization to determine the RM. The two methods are
complementary because the $Q-U$ fitting method has a solid statistical foundation in parameter inference, while the RM synthesis method
provides a global view of the linear polarization intensity as a
function of the wide RM range\cite{SL17}. As a
parameter estimation problem, it is preferential to use $Q-U$ fitting with
Bayesian inferences\cite{DKL19} as the primary method to measure
the Faraday rotation in the observer frame. Nevertheless, we compare the result obtained using this method 
with that obtained using the RM synthesis method\cite{rmsyn15} to cross-check the measured RM values and to make sure that the
inferred parameters do not fall into a local minimum. 

The Bayesian method\cite{DKL19} includes modelling the systematics and  
deviation of RM-induced linear polarization wrapping, so it generally produces 
reliable RM values. The Stokes parameters $Q$ and $U$ are integrated in the 
burst time intervals, binned to a frequency resolution of 
$1.95\,\mathrm{MHz}$ per channel. As shown in \EXTFIG{fig:qu}, we detected clear RM features for each burst, i.e. the sinusoidal curves of the Stokes 
parameters $Q$ and $U$ as functions of wavelength. After the RM measurement, we used the software package \textsc{IonFR}\cite{SSH13} to compute the ionospheric corrections, where values are bellow $1\,\radm$. The corrected RM values are listed in \TAB{tab:rpts}. 

We cross-checked the RM values using the 'revisited' RM synthesis\cite{rmsyn15}.
The polarization dynamic spectra were normalized by the baseline r.m.s.
value of the total intensity in each channel. Then, the Stokes parameters
$Q$ and $U$ were derotated with trial RM values ranging from $-8000\,\radm$
to $+8000\,\radm$, a range chosen to cover much more than the
original RM of ${-3163\pm 20\,\radm}$ given in the initial
discovery\cite{Price19MN}. Here, the 'revisited' derotation vector
was used to correct the systematics caused by uniform frequency
sampling, as explained in ref.\cite{rmsyn15}. The degree
of linear polarization for each trial RM was then calculated from
the linearly polarized profile and the total intensity profile.
Following the convention\cite{HvSM04}, the uncertainty of the RM
values was estimated using half of the FWHM divided by the S/N value of the 
local Gaussian peak. The RM spectra are shown in \EXTFIG{fig:rmsyn_bursts}.
The measured RM values are compared with the Bayesian values
in \TAB{tab:rpts} and \FIG{fig:rmvar}a.

We checked whether the RM value or polarization profile could be affected
by the value of DM used in dedispersion by comparing the 
measured RM values produced by a different data-processing pipeline, 
in which we fixed the DM for all bursts. We chose to fix the DM to the value of burst 5, which has the widest bandwidth and the highest S/N among all 
bursts. 
The measured RM values for aligned/fixed-DM
sets agree with each other within the uncertainties.

To make sure that the RM measurement was not affected by the polarimetry
instability of FAST or off-axis illumination, we conducted three
extra test observations with a pulsar, PSR J1915+1009, on 16 and 17 January 2020.
The DM and RM values of PSR J1915+1009 are similar to
those of FRB 180301, and its position is known accurately. In the first test
observation, the pulsar was placed at the centre of the central
beam; in the second test observation, the pulsar was placed $2.6'$ 
(see Methods section 'Localisation and position accuracy') away from
the beam centre; and the third test observation was conducted one
day later with the pulsar at the beam centre. We checked the temporal
stability of the Bayesian RM measurement using the first and third test observations. As shown in \EXTFIG{fig:poln_test}a, 
no RM variation is found in the test pulsar observation. The maximum RM
variation is $0.1\,\radm$ over the two-day timescale.  We
examined the off-axial polarization error by comparing the RM measured
in the first and the second observation sessions. As shown in \EXTFIG{fig:poln_test}b, the maximum difference is
$0.3\,\radm$. We also checked the linear and circular degrees 
of polarization by comparing the pulse profiles of the three observation 
sessions carried out at FAST with previously published results\cite{JK18}. 
As one can see in \EXTFIG{fig:poln_test}c-e, both linear and circular 
polarizations are stable. Therefore, the RMs that we measured for FRB 180301 are not affected by instrumental instability.

Our RM values are highly different from that reported 
(${-3163\pm 20\,\radm}$) in the initial
discovery\cite{Price19MN}, in which the signal was only bright in
a 40-MHz frequency window and the so-produced linear polarization in the
integrated profile was not apparent. Although
it is possible that the RM value changed so dramatically owing to the
compact magnetoionic environment, such a discrepancy in RM values
probably comes from the narrow-band bias of the RM transfer
function\cite{2015MNRAS.450.3579S}, that is, the previously reported value
may be biased because of the just 40~MHz effective bandwidth.
With the RM values that we obtained, the degree of linear polarization for 
the RM-corrected pulse profile is rather high ($36-80\%$). 

Such a high RM of $\sim550\,\radm$ cannot
be easily explained by the Galactic foreground, which is measured to be $+ 72 \pm 8\,\radm$ at 
these Galactic coordinates, according to Xu \& Han\cite{XH15RAA} (see also the 
web link \url{http://zmtt.bao.ac.cn/RM/searchGRM.html}). Likewise, the intergalactic RM contribution\cite{XuHan14} is estimated to be at most a few to a few tens of radians per square metre. The
remaining observed RM is contributed by the host galaxy or local environment
at redshift $z \sim 0.13-0.35$, which implies that the intrinsic RM in the source rest frame is $({\rm RM_{obs}-RM_{Gal}}) \times (1+z)^2 = (550-72) \times (1.27\ {\rm to}\ 1.82) \sim 600\ {\rm to}\ 870\,\radm$. 
The most probable origin of this large RM is from
the local environment$-$as opposed to the interstellar medium of the
host galaxy.

As one can see, the RM values measured with the Bayesian method and
the RM synthesis are consistent. For six out of seven measurements, 
the results of the two methods agree within 1$\sigma$,
and all seven measurements agree within 2$\sigma$. Because
the two methods belong to different classes of statistics, their
errors have different meanings. Nevertheless, we note that the RM synthesis 
tends to produce larger errors than the Bayesian method. We also 
combine the RM synthesis spectra of different bursts to fit the RM values 
jointly. The combined RM spectra are shown in \EXTFIG{fig:rmsyn_global}. The
jointly fitted RM values are $535.8\pm 2.7\,\radm$
on 6 October 2019 (bursts 5, 7, 9, 10), $558.9\pm
3.7\,\radm$ on 7 October 2019 (bursts 11, 12, 13), and
$543.7\pm 2.6\,\radm$ for all seven bursts.
The three joint-fitting results indicate an RM variation. Using the Bayesian 
RM values in \TAB{tab:rpts}, the computed $\chi^2$ becomes $\chi^2=450$ 
(corresponding to 21$\sigma$). If we use RM synthesis values, $\chi^2=51\times6\sigma$. 
The RM values for data dedispersed with the DM of burst 5 produces a similar level of $\chi^2$, and the conclusion does not change. 
The $\chi^2$ test indicates that the RM is probably not a constant.

We investigated whether the RM changes during the bursts
but find no such variation. Phase-resolved RM variation has been reported f
or radio pulsars\cite{NKK09} and for FRB 181112\cite{Cho20}. 
As shown by Noutsos et al.\cite{NKK09}, the phase-dependent RM variation may 
come from scattering in interstellar ionized medium or from 
propagation effects in the magnetosphere. Cho et al.\cite{Cho20} noted that 
the RM differs by approximately 25 $\radm$ within one burst from FRB 181112. 
Here, the null detection of phase-dependent RM variation in FRB 180301 could 
be a consequence of the limited S/N. However, it is also possible that the 
scattering effects of FRB 180301 is too weak to produce RM variation during the burst.

We checked whether the measured RM spectra are consistent with the thin-screen model. Data of burst 5 were used for the test because it has the best S/N. As shown in \EXTFIG{fig:rmspeccmp}, after we subtract the linear polarization contribution of the best-fitted RM (that is, the spectrum of a thin Faraday screen), the remaining RM spectrum is consistent with noise, suggesting that the radiation comes from a compact region with the same Faraday depth. 

\subsubsection*{Polarization angles of bursts}

After correcting for the RM, we have the infinite-frequency polarization
profile of each burst. The RM-corrected profiles with the individual
Bayesian RM values are shown in \FIG{fig:poln2}. For the
diverse PA swings observed from FRB 180301, even though bursts 5, 9, 11 and 13 
may be characterized as pulsar-like shapes, it is unlikely that a simple rotating-vector model invoking
a dipolar magnetic field\cite{RadhakrishnanApL1969} can reproduce the properties of all the bursts.

We note that there are obvious differences between the centroids
of the PAs, if we derotate the linear polarization with
a globally fitted RM (as shown in \EXTFIG{fig:global_pa}). If we derotate
the linear polarization with the individual RM values in \TAB{tab:rpts}, 
we can approximately bring the centroids of
the PAs to a constant value after taking the RM errors into account. 
Thus, either there is RM variation in FRB 180301 or the central values of 
PA swings vary pulse-to-pulse. During the derotation of linear
polarization with different RMs, the PA swing of each burst keeps
its shapes, as seen from comparing \EXTFIG{fig:global_pa} and 
\FIG{fig:poln2}.

The shape of PA swing may depend on the DM. For example, if the DM value is inaccurate, the polarization features will be smeared and the PA swing will be flatter. 
To test whether the diversity of PA swing in \FIG{fig:poln2} is affected by DM, we compare it with the PA swings produced using the maximum, minimum and burst-5 DM values. The polarization profiles of all cases 
are presented in \EXTFIG{fig:poln2_dm5}. There is little change in the shape of PA for all the bursts, even though bursts 7, 10 and 12 have a relative shift in the PA centroid. Since every 3 $\radm$ of RM change leads to approximately 20$^\circ$ of parallel shift in PA for the central frequency at 1250 MHz, the PA shifts for bursts 7, 10, and 12 are due to a sub-error change of the RM values, as discussed in Methods section 'polarization and Faraday rotation'. 

\subsubsection*{Coherent radiation models}
The extremely high brightness temperatures as observed in FRBs 
require that the radiation mechanism must be coherent. In the literature, two broad classes of models\cite{Lu18MN} have been discussed: pulsar-like models invoking magnetospheres\cite{Katz14PRD,Kumar17MN,Zhang17ApJL,YangZhang18ApJ} and 
$\gamma$-ray-burst-like models\cite{Lyubarsky14MN,Waxman17ApJ,Plotnikov19MN} considering relativistic shocks. 
The latter invoke synchrotron masers as the coherent mechanism to emit FRBs. There are two sub-types of synchrotron maser models. 
One type\cite{Waxman17ApJ} considers a non-magnetized shock, so that 
emission is not expected to be highly polarized. The other 
type\cite{Lyubarsky14MN,Metzger19MN,Beloborodov19,Plotnikov19MN} invokes highly ordered magnetic fields in the shock plane to allow particles to emit coherently; to achieve this, this model requires that the magnetic field lines in the upstream medium are highly ordered. As a result, such a model predicts a constant PA across a single burst\cite{Metzger19MN,Beloborodov19}. 
To account for variation of PA across a burst, this model needs to introduce a variation of magnetic field configuration as a function of radius in the upstream medium through which the shock propagates. A striped wind from a rotating pulsar may in principle provide a varying magnetic field configuration in the shock upstream as a function of radius\cite{Sironi11}. However, this model can only generate monotonic, smoothly varying PA curves and may not account for the diverse PA 
evolution patterns presented in \FIG{fig:rmvar}, such as non-monotonic variations (for example, bursts 7 and 12) and jumps in PA between sub-pulses (for example, bursts 10, 11 and 13). More complicated upstream magnetic field configurations may be possible if the medium has undergone violent disturbance due to a previous shock. However, in this case, the fields are no longer ordered, and the maser mechanism cannot operate. On the other hand, diverse PA variation features have been frequently observed in radio pulsars and flaring magnetars, the radio emission of which is believed to originate from the magnetosphere of a neutron star\cite{Manchester77,Camilo16ApJ}.

Magnetospheric models attribute PA variations to line-of-sight sweeping across the radiation beam. These models include the standard scenario invoking neutron star rotation (as is the case of radio pulsars and flaring magnetars)\cite{RadhakrishnanApL1969,Manchester77} and a scenario invoking sudden reconfiguration of a magnetosphere triggered by an external ram pressure\cite{Zhang17ApJL}. Propagation effects$-$either inside the magnetosphere or far from the source, outside the magnetosphere$-$may provide additional mechanisms to introduce PA variation. However, these effects tend to introduce a systematic effect on all the bursts, unless the environments (for example, plasma lensing) vary rapidly within timescales of hours, which is unlikely\cite{Cordes19ARAA}. The fact that the PA variation patterns differ for different bursts of FRB 180301 suggests that propagation effects do not play the dominant role in shaping the PA variation patterns. Whether or not a PA variation is observed and how PA varies with time depend on the magnetic field configuration and the line-of-sight geometry. For example, the simple rotating-vector model for a dipolar magnetic field configuration predicts an 'S-' or 'inverse-S-' shape in PA variation curves. More complex magnetic field configurations can produce more complicated PA variation curves. Nearly straight magnetic field lines sweeping the line-of-sight can produce non-varying PA patterns. This may be realized either in a rotating neutron star model with the emission region in the outer magnetosphere, or in the interaction model. The diverse PA swings observed from FRB 180301 and the failure of the simple dipolar rotating-vector model\cite{RadhakrishnanApL1969} suggest that the system may have complicated magnetic field configurations and line-of-sight geometries, which may vary with time. Such a scenario may be achieved in interacting models\cite{Zhang20} rather than simple rotating models. 
This is also consistent with the lack of strict periodicity observed in repeating bursts\cite{Zhangy18}. Within the magnetospheric model, the fact that many FRB bursts (including those from other sources, 
such as FRB 121102\cite{Michilli18Nat} and FRB 180916.J0158+65\cite{CHIME19ApJL})
showed a nearly constant PA during each burst would suggest that their emission region is in the outer part of the magnetosphere where the field lines are nearly straight.

Both FRB 180301 (with varying PA) and FRB 121102 (with non-varying PA) show similar emission properties, such as complex pulse profiles\cite{Spitler16Nat} and downward-frequency-drifting sub-pulses\cite{Hessels19ApJL}. These features can be understood within a generic magnetospheric model using open-field-line regions as the FRB emission site\cite{Wang19ApJL,Lyutikov20}. As the emission unit moves away along the open-field-line regions (either for the polar-cap geometry or the cosmic-comb geometry), an observer always sees higher-frequency emission earlier than lower-frequency emission\cite{Wang19ApJL}.

\subsubsection*{Burst rate}
Using the Kolmogorov-Smirnov test, we we compare the Weibull, Gaussian and exponential distributions and find that the one that best describes the burst waiting time is the Weibull function\cite{Oppermann18MN}.The p-values for each case are 0.728, 0.004 and 0.202, respectively, so we 
can use either the Weibull or the exponential distribution to describe the waiting times. Here we adopt a Weibull distribution because the inference on its shape parameter can help to understand if there is significant deviation from a Poisson process. 

The likelihood function of the waiting time series $\Delta T-$that is, its probability density function$-$is 
\begin{equation}
    \mathcal{W}(\Delta T|k,r)=\frac{k}{\Delta T}\left[r \Delta T \Gamma\left(1+\frac{1}{k}\right)\right]^{k}e^{-\left[r \Delta T \Gamma(1+\frac{1}{k})\right]^k}\,,
\end{equation}
with the $\Gamma$ function defined as $\Gamma(x)=\int^{\infty}_{0} t^{x-1}e^{-t}\D t\,$.
Here, $r$ is the burst rate and $k$ is the shape parameter. With the above 
likelihood function and assuming uniform priors, we carried out a Bayesian 
inference using the nested sampling software package 
\textsc{MultiNest}\cite{Feroz09MN}. According to the data from all 
the observational sessions in \TAB{tab:rpts} and \EXTTAB{tab:obs}, and the peak flux density threshold of FAST 
at $\simeq 5\,\rm mJy$ (see \TAB{tab:rpts}), the parameters are inferred as 
$r=1.2^{+0.8}_{-0.7}\,\rm hr^{-1}$ and $k=0.9^{+0.3}_{-0.3}$ with a 95\% 
confidence level. The posterior is plotted in \EXTFIG{fig:weib}. The shape parameter is close to 1, which suggests that the distribution can also be well described by a Poisson process.

\clearpage
\noindent {\bf References} \\

\clearpage

\subsubsection*{Data availability}

The data that support the findings of this study are available at   
\url{https://psr.pku.edu.cn/index.php/publications/frb180301/}. 

\subsubsection*{Code availability}
The BEAR package is available at \url{https://psr.pku.edu.cn/index.php/publications/frb180301/}


\begin{addendum}
 \item This work used data from FAST, a Chinese
 national mega-science facility, built and operated by the National
 Astronomical Observatories, Chinese Academy of Sciences. This work
 is supported by the Natural Science Foundation of China (U15311243,
 11988101, 11833009, CAS XDB23010200), the Cultivation Project for FAST
 Scientific Payoff and Research Achievement of CAMS-CAS, the
 Max-Planck Partner Group, NKRDPC 2017YFA0402600, the Youth Innovation
 Promotion Association of CAS (2018075).

 \item[Author Contributions] R.L. led the observational proposal
 2019a-129-P in the FAST 'Shared-Risk' observations and the statistical
	 analysis of repeating events. B.J.W., Y.P.M., C.F.Z., and K.J.L.
	 developed the searching pipeline and processed the raw
	 data to produce FRB candidates. J.C.J. conducted the polarization
	 calibration and RM measurements. H.X.
	 conducted the flux calibration. R.N.C.
	 and Y.J.G. performed the timing analysis. B.Z.,
	 W.Y.W., R.X.X. and J.P. provided theoretical discussions. J.Y., M.W.
	 and N.W. contributed to discussions on observation planning. K.J.L.,
	 J.L.H., and B.Z. organized the FRB searching team, co-supervised
	 the data analysis and interpretations and led the writing of the paper.
	 The search software BEAR was tested by M.Z.C., X.L.C., L.F.H., Y.X.H., J.L., Z.X.L., J.T.L., X.P., Z.G.W. and
	 Y.H.X. FAST observations, instrument setting and monitoring
	 was done by P.J., L.Q., H.Q.G., H.L., J.H.S., J.Y., D.J.Y. and Y.Z..  
	 All authors contributed to the analysis or interpretation of
	 the data and to the final version of the manuscript.

 \item[Competing Interests] The authors declare no
competing interests.

\item[Correspondence and requests for materials] should be addressed to K.J.L., J.L.H. or B.Z.

\end{addendum}
\clearpage

\section*{Extended Data}

\renewcommand{\baselinestretch}{1.0}
\selectfont

\setcounter{figure}{0}
\setcounter{table}{0}

\captionsetup[table]{name={\bf Extended Data Table}}
\captionsetup[figure]{name={\bf Extended Data Fig.}}

\begin{table*}[!htbp] 
    \centering
		\caption{\textbf{{FAST observations of FRB 180301}}}
    \begin{tabular}{cccc}
    \hline
       Observational Time & MJD & Duration & Number of \\
         (UTC) & & (hr:min) & detections \\
    \hline
      2019-07-16 02:00:00 - 2019-07-16 04:00:00 & 58680 & 2:00 & 4 \\
      2019-09-11 21:52:00 - 2019-09-12 01:48:00 & 58737 & 3:56 & 0 \\
      2019-10-06 20:49:09 - 2019-10-06 23:49:09 & 58762 & 3:00 & 6 \\
      2019-10-07 20:33:23 - 2019-10-07 23:27:23 & 58763 & 2:56 & 5 \\
    \hline
    \end{tabular}
    \label{tab:obs}
\end{table*}

\clearpage

\begin{figure}  %
\centering
\includegraphics[width=180mm]{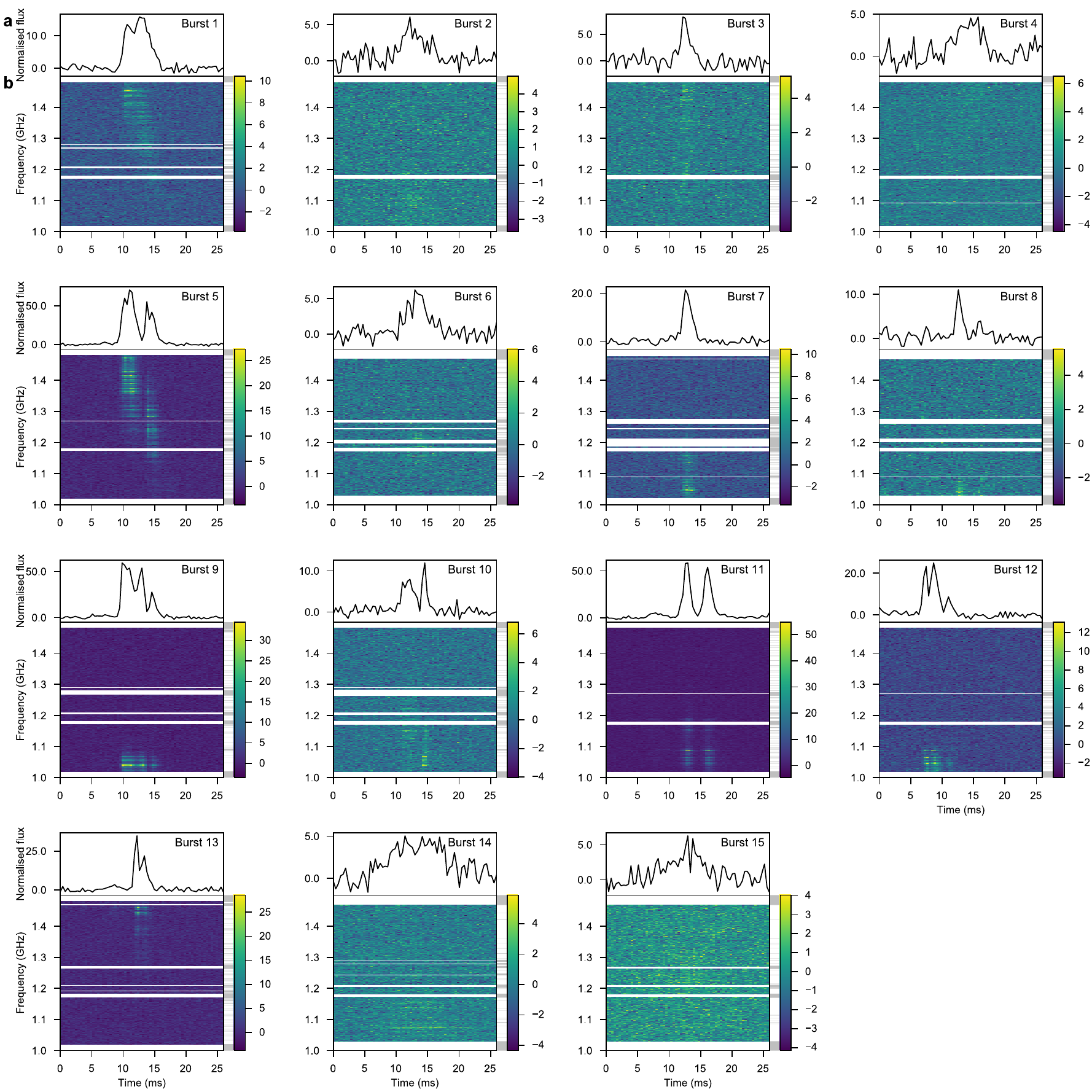}
\caption{\textbf{Dynamic spectra for all 15 detected bursts of FRB 180301.} 
{\bf a}, Dedispersed pulse profile. {\bf b}, Dynamic spectra for the 
total intensity as a function of frequency and time (with a 
frequency resolution $1.95\,\mathrm{MHz}$ per channel and a time resolution 
of $393.2\,\mathrm{\mu s}$ per bin). The colour bars denote the intensity 
S/N scaled with the off-pulse r.m.s. value. The $\DM_{\rm aligned}$ in 
\TAB{tab:rpts} is used to dedisperse each burst. For bursts 1-4, we plot the 
raw intensity because only two linear polarization channels were recorded. 
For the rest of the bursts, polarization calibrations were performed.
\label{fig:all_bursts}}
\end{figure}

\clearpage

\begin{figure}  
\centering
\includegraphics[width=180mm]{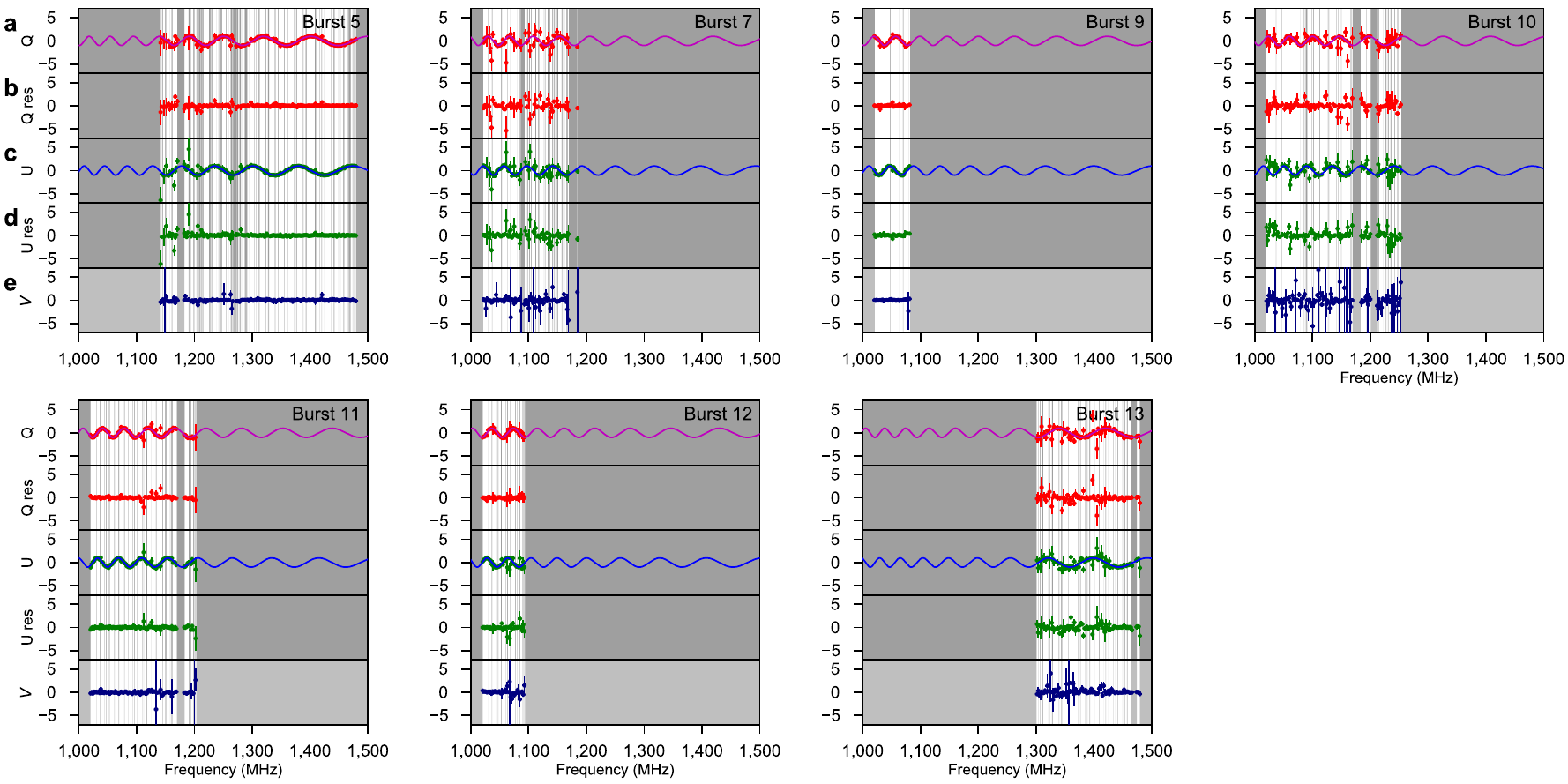}
\caption{\textbf{Observed and fitted Stokes parameters $Q$ and $U$ for 
linear polarization as a function of frequency.} {\bf a}, {\bf c}, 
Normalized Stokes parameter $Q$ and fitting residuals.  {\bf b}, {\bf d}, 
normalized Stokes parameter $U$ and fitting residuals. The amplitudes of 
the oscillation have been normalized using the inferred linear polarization 
intensity. 
{\bf e} Stokes parameter $V$ normalized by total intensities in each channel. 
The grey shaded frequencies are removed before fitting owing to low signal 
intensities, RFI or band-edge effects. The error bars denote the 68\% confidence intervals. 
The burst number in each subplot is as in \TAB{tab:rpts}.
\label{fig:qu}}
\end{figure}

\begin{figure} 
    \centering
		\includegraphics[width=180mm]{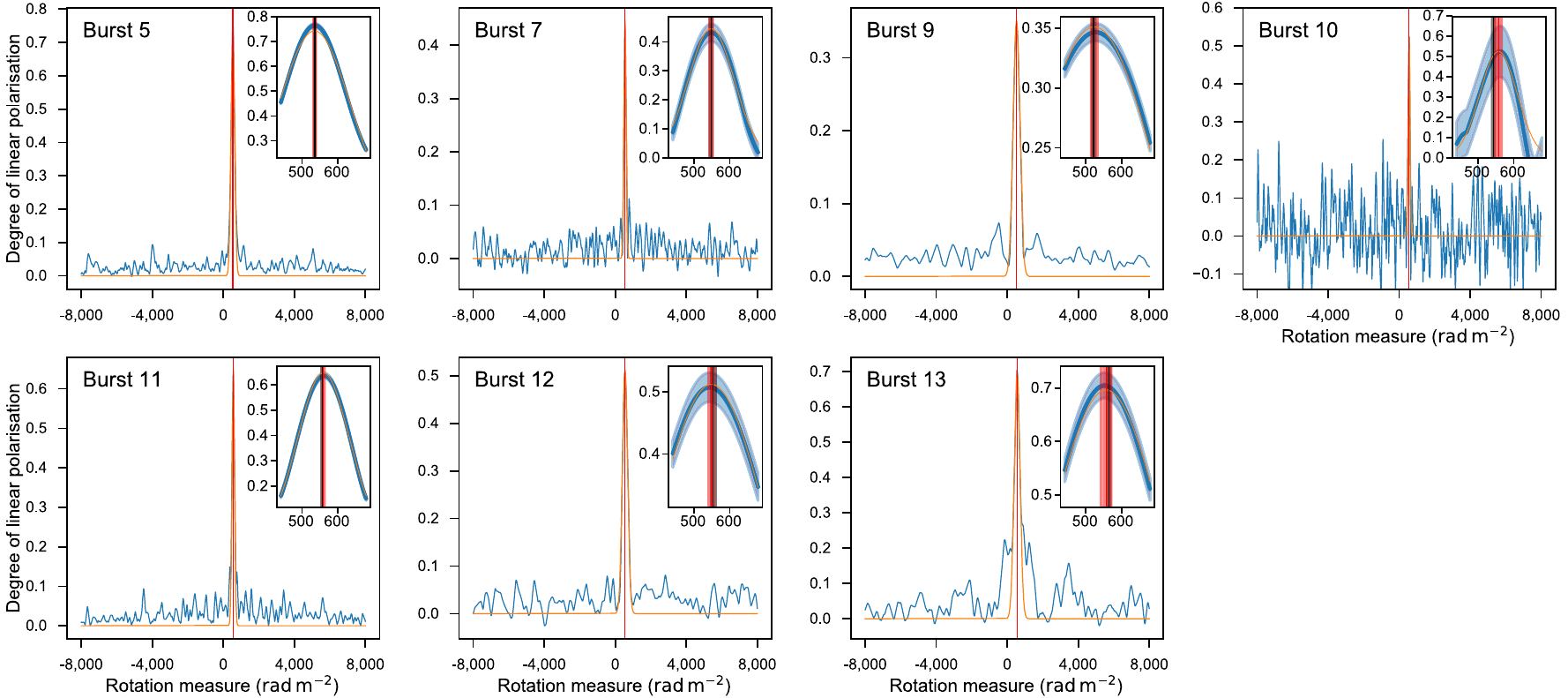}
		\caption{\textbf{RM synthesis results of the seven bursts}. 
		We calculate the RM spectrum within the range $-8000$ to 
		$+8000\,\radm$. The horizontal red shaded area denotes the 
		1$\sigma$ interval of the baseline. The vertical red line denotes the best-fit RM value. We also show a zoom-in of the spectral peak, where the vertical orange dashed lines show the range in which the spectrum is used in peak fitting. The best-fitting Gaussian and its 68\% confidence interval are 
		indicated by the orange curve and blue shading. The vertical red lines and shading show the best-fit RM and the 68\% confidence intervals. We also show the Bayesian RM here, indicated by the vertical black lines and shading. The burst number in each sub-plot is defined in \TAB{tab:rpts}.
		\label{fig:rmsyn_bursts}}
\end{figure}

\begin{figure} 
    \centering
		\includegraphics[width=180mm]{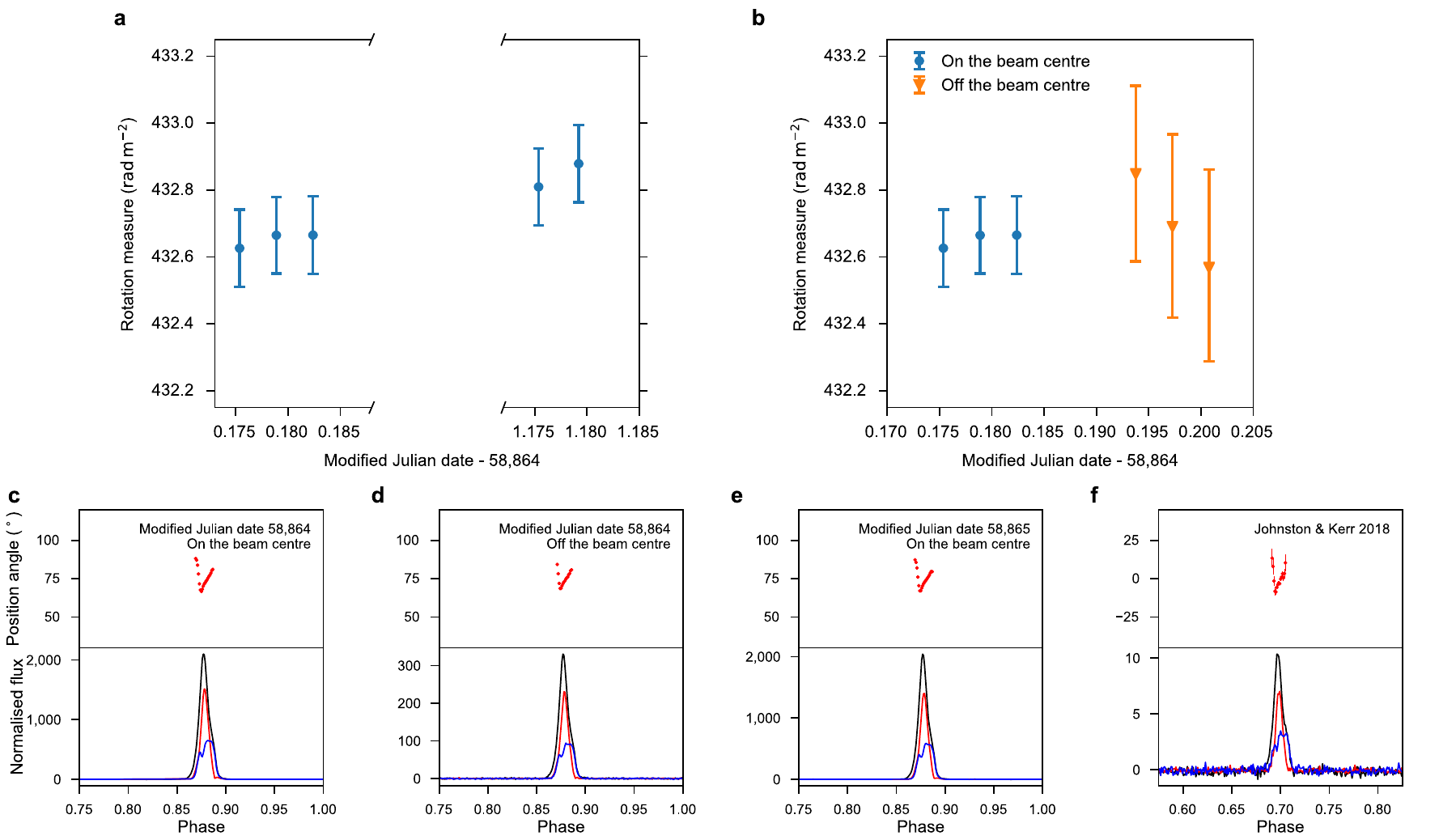}
		\caption{\textbf{Polarimetry stability test}. {\bf a}, Temporal 
		stability test. The RM values of PSR J1915+1009 measured with the Bayesian method confirm 
		that there is no obvious RM variation in one-day interval. The error bars
		denote 68\% confidence intervals. {\bf b} Off-axis polarimetry test.  
		PSR J1915+1009 was first placed in the beam centre and then $2.6'$
		away from the beam centre. The RM values measured with the Bayesian method 
		confirms that there is no apparent systematic error for the off-axis 
		illumination. The off-axis data point has a larger error because S/N drops 
		for those observations owing to the off-axis illumination.  
		{\bf c}, Polarization pulse profile and PA for PSR 
		J1915+1009, observed with central illumination. {\bf d}, As in {\bf c}, but off-axis illumination is used. {\bf e}, Polarization profile and PA with central illumination observed one day later. {\bf f}, polarization pulse profile measured using the Parkes radio telescope by Johnston and Kerr\cite{JK18}.
		\label{fig:poln_test}}

\end{figure}

\begin{figure} 
    \centering
		\includegraphics[width=135mm]{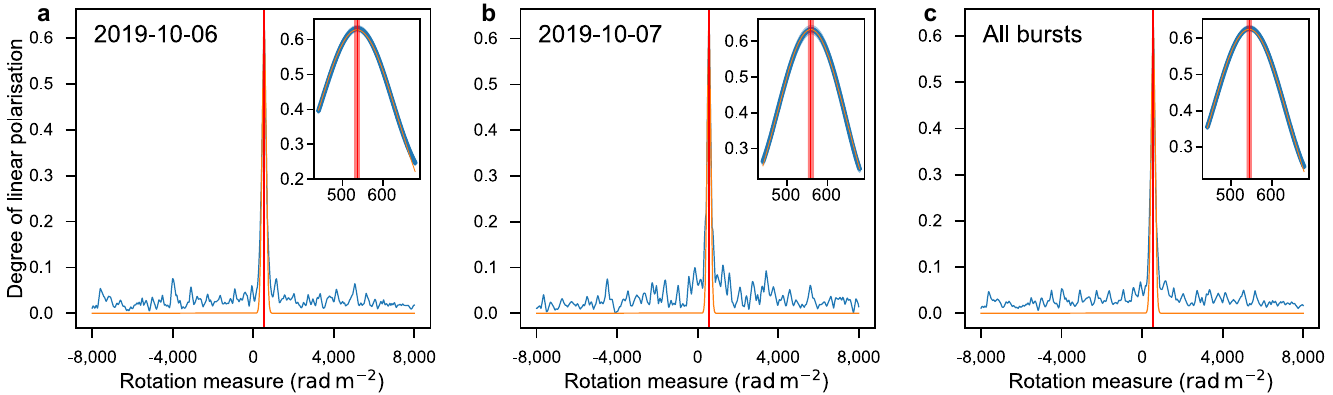}
		\caption{\textbf{The joint fitting results of the RM synthesis spectra}.  
		{\bf a}, Burst 5, 7, 9, 10 on 6 October 2019. {\bf b}, Burst 11, 12, 13 on 7 October 2019.
		{\bf c}, All seven bursts. The notation is the same as 
		\EXTFIG{fig:rmsyn_bursts}.
		\label{fig:rmsyn_global}}
\end{figure}
\clearpage

\begin{figure} 
    \centering
		\includegraphics[width=89mm]{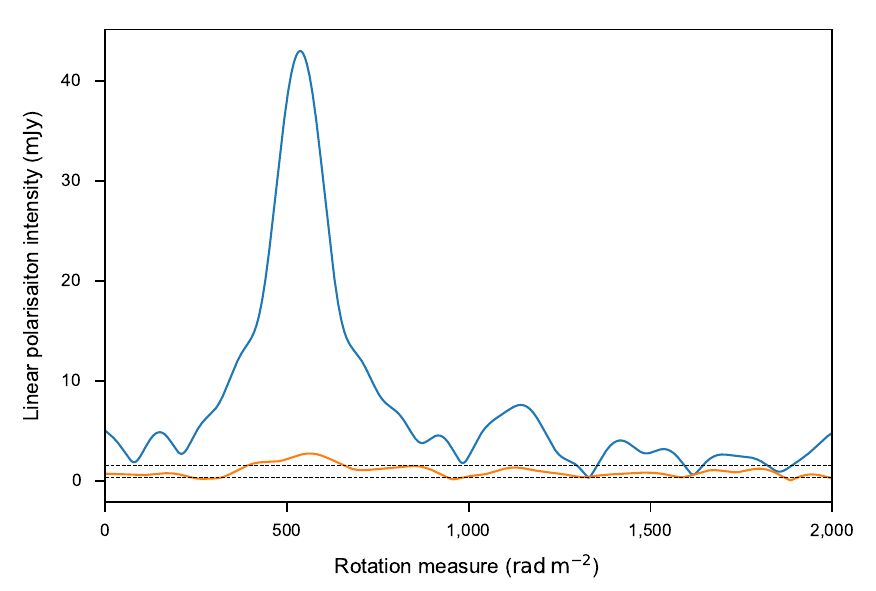}
		\caption{\textbf{RM synthesis spectra before and after thin screen subtraction.} 
		The blue and orange curves are the RM synthesis intensity spectra for 
		burst 5. The orange curve is computed after subtracting the Stokes 
		$Q$ and $U$ corresponding to the RM of burst 5. The orange curve is 
		consistent with noise. This indicates a thin-screen scenario for the Faraday rotation.
		\label{fig:rmspeccmp}}
\end{figure}
\clearpage

\begin{figure} 
    \centering
		\includegraphics[width=180mm]{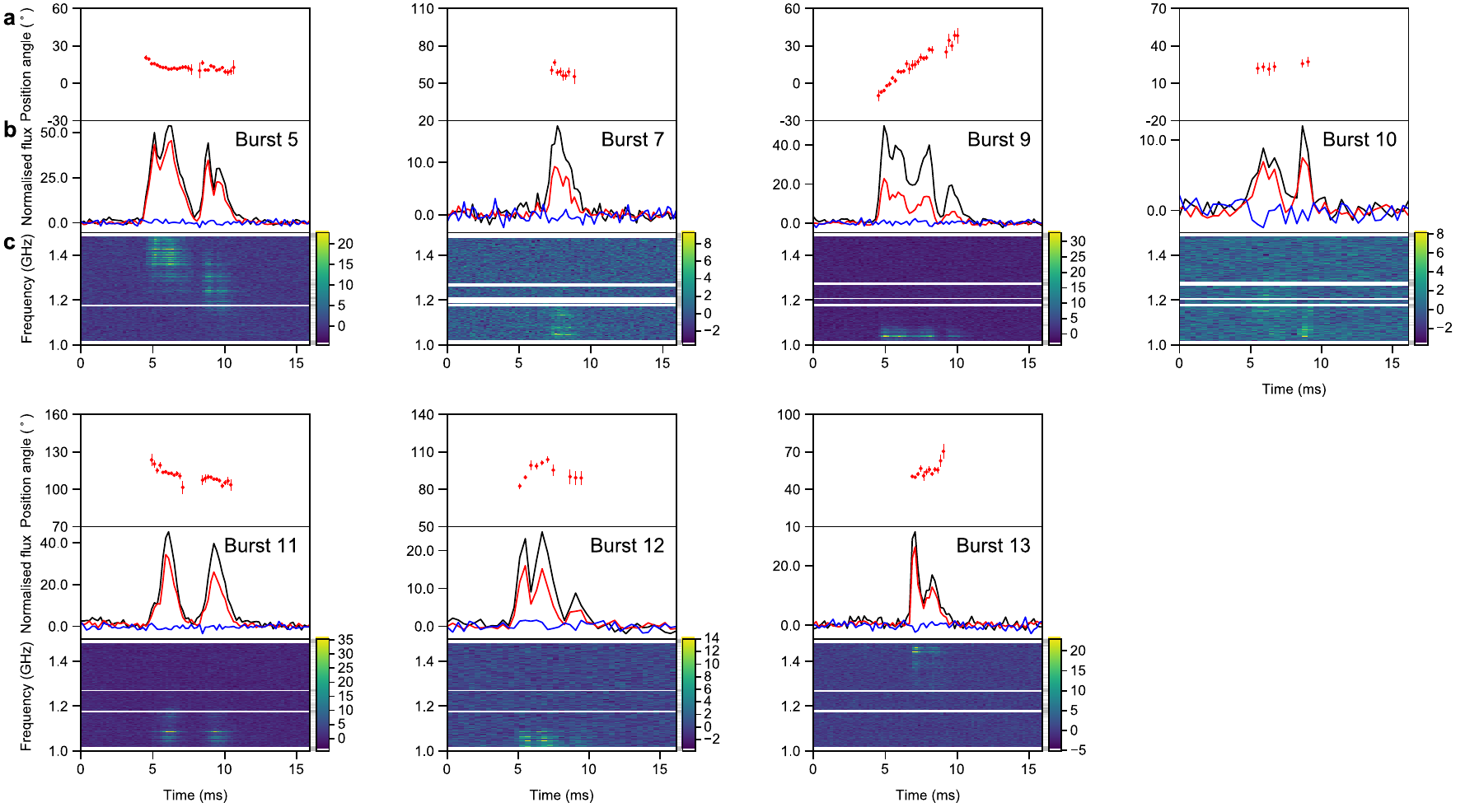}
		\caption{\textbf{Polarization profiles of seven bright bursts 
		and their dynamic spectra.} Here we used the globally fitted {${\rm 
		RM}=543.7\pm 2.6\,\radm$} to derotate the linear polarization. 
		The other settings are the same as in \FIG{fig:poln2}.
		\label{fig:global_pa}}
\end{figure}

\clearpage

\begin{figure} 
\centering
\includegraphics[width=180mm]{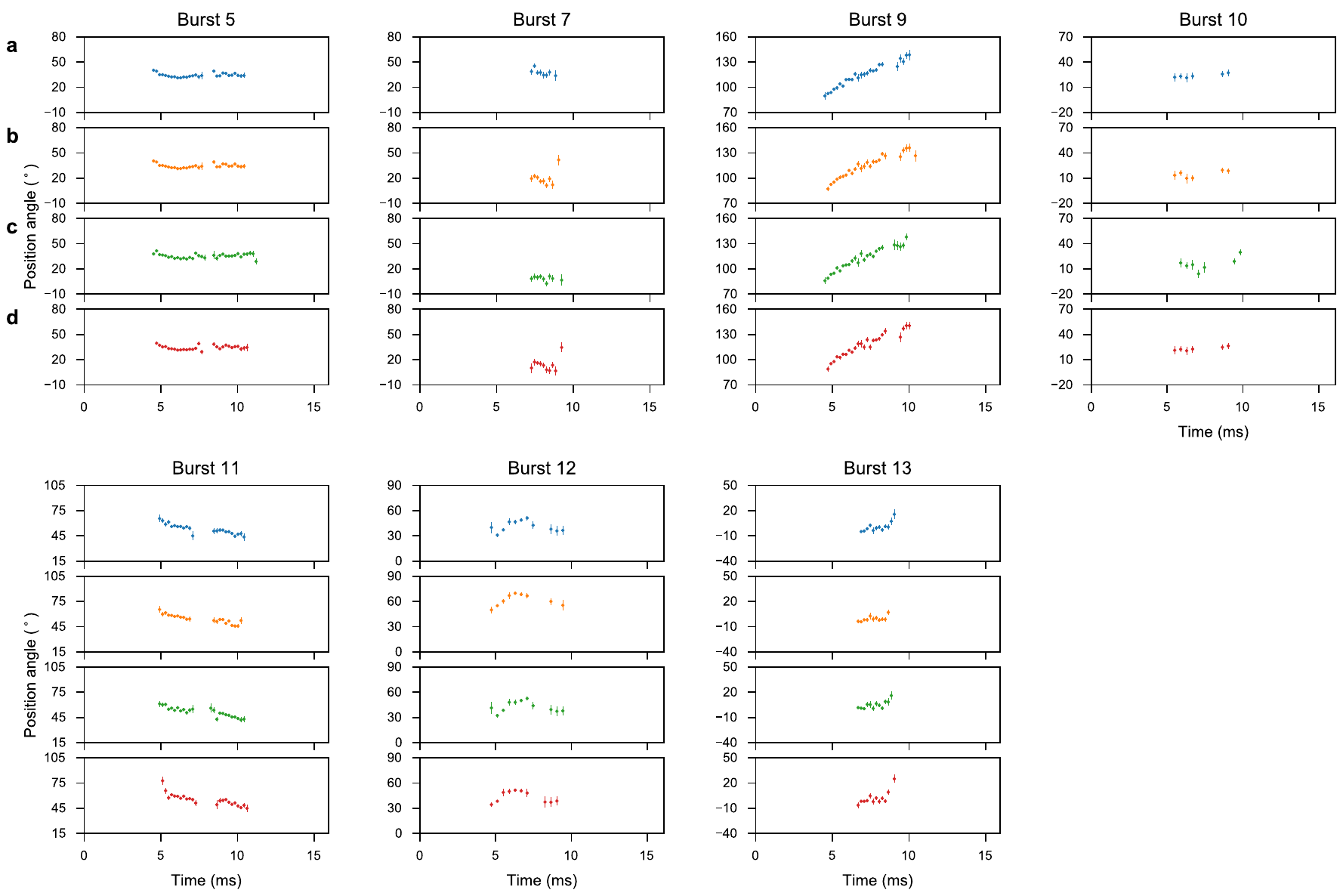}
\caption{{\textbf{Comparison of PA swing from seven bright bursts using different DM values in dedispersion.}} {\bf a}$-${\bf d}, For each burst, blue curves use individually measured DM values as in \TAB{tab:rpts} ({\bf a}), orange curves use the DM of burst 5 ({\bf b}),  green curves use the lowest DM (from burst 12) ({\bf c}) and red curves use the highest DM (from Burst 10) ({\bf d}) .
\label{fig:poln2_dm5}}
\end{figure}

\clearpage

\begin{figure}  
\centering
\includegraphics[width=180mm]{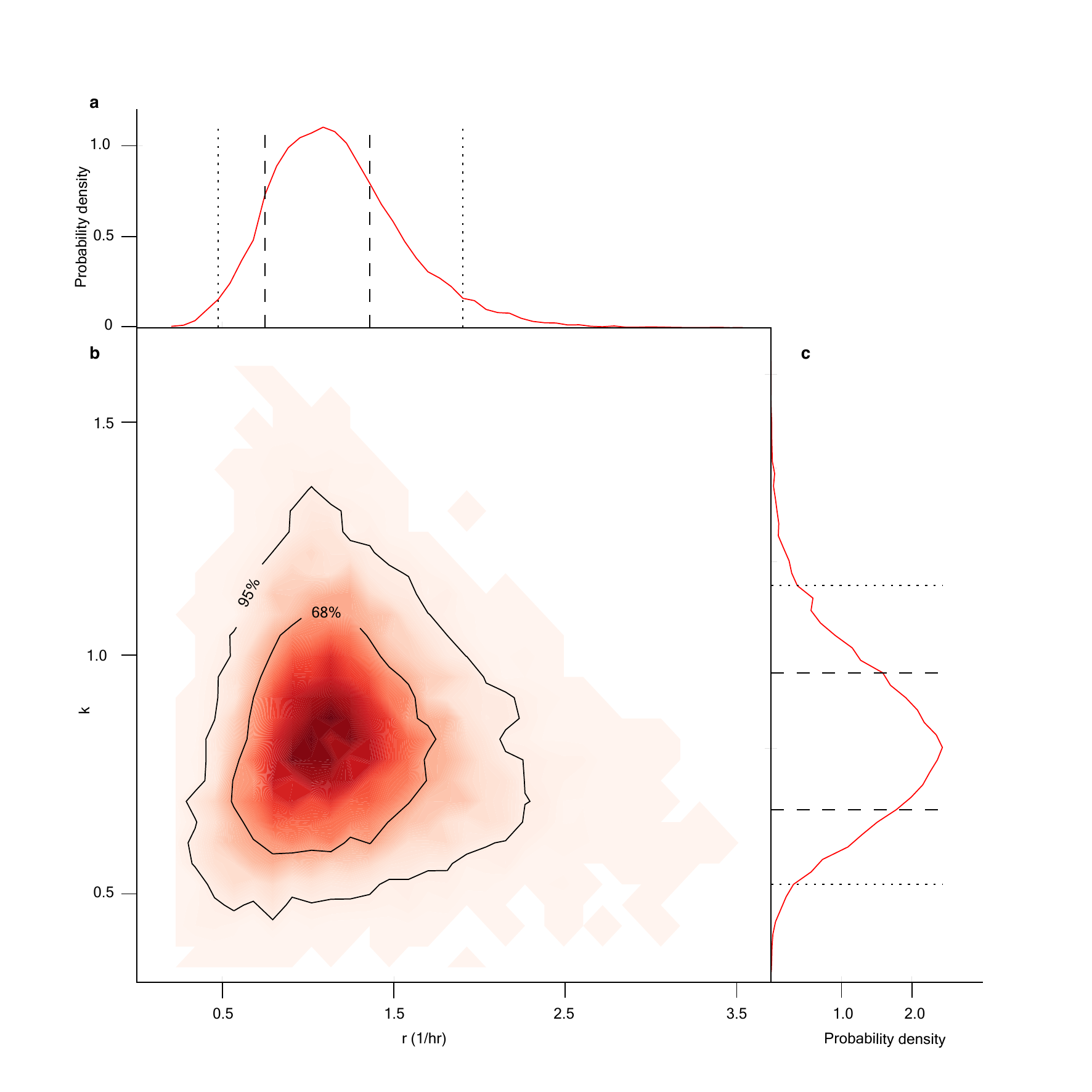}
\caption{\textbf{Posterior distribution for the burst rate inference.} 
\textbf{a}, Marginalized posterior of the burst rate. The dashed and dotted 
lines denote 68\% and 95\% confidence levels, respectively. \textbf{b}, Two-dimensional distribution of the posterior. The horizontal and vertical axes show the burst rate and the shape parameter of the Weibull distribution, respectively. \textbf{c}, Marginalized posterior for the shape parameter.
\label{fig:weib}}
\end{figure}



\end{document}